\documentclass[3p,letterpaper]{elsarticle}

\usepackage{graphicx}
\usepackage{bm}
\usepackage{hyperref}

\begin{document}

\begin{frontmatter}

\title{The End of Hyperbolic Growth in Human Population and CO$_2$ Emissions\tnoteref{t1}}
\tnotetext[t1]{Published in \textit{Physica A} {\bf 661}, 130412 (2025), \url{https://doi.org/10.1016/j.physa.2025.130412}}

\author{Victor M.~Yakovenko}
\ead{yakovenk@umd.edu}

\affiliation{organization={Department of Physics and JQI, University of Maryland}, city={College Park}, addressline={Maryland}, postcode={20742}, country={USA}}

\date{14 August 2025}

\begin{abstract}

Using current empirical data from 10,000 BCE to 2023 CE, we re-examine a hyperbolic pattern of human population growth, which was identified by von Foerster {\it et al.}\ in 1960 with a predicted singularity in 2026.  We find that human population initially grew exponentially in time as $N(t)\propto e^{t/T}$ with $T=2080$ years.  This growth then gradually evolved to be super-exponential with a form similar to the Bose function in statistical physics. Around 1700, population growth further accelerated, entering the hyperbolic regime as $N(t)\propto(t_s-t)^{-1}$ with the extrapolated singularity year $t_s=2030$, which is close to the prediction by von Foerster {\it et al.}  We attribute the switch from the super-exponential to the hyperbolic regime to the onset of the Industrial Revolution and the transition to massive use of fossil fuels.  This claim is supported by a linear relation that we find between the increase in the atmospheric CO$_2$ level and population from 1700 to 2000.  In the 21st century, we observe that the inverse population curve $1/N(t)$ deviates from a straight line and follows a pattern of ``avoided crossing'' described by the square root of the Lorentzian function.  Thus, instead of a singularity, we predict a peak in human population at $t_s=2030$ of the time width $\tau=32$ years.  We also find that the increase in CO$_2$ level since 1700 is well fitted by ${\rm arccot}[(t_s-t)/\tau_F]$ with $\tau_F$=40 years, which implies a peak in the annual CO$_2$ emissions at the same year $t_s=2030$.

\end{abstract}


\end{frontmatter}

\section{Introduction}
\label{Sec:Intro}

In 1960, Heinz von Foerster {\it et al.}\ published a paper \cite{vonFoerster-1960} with the provocative title ``Doomsday: Friday, 13 November, A.D.\ 2026.''  Fitting the data then available, the paper argued that growth of human population is described by a hyperbolic function of time that extrapolates to infinity in the year 2026.  The particular day of 13 November was a joke, as it is the birthday of Heinz von Foerster.\footnote{\url{https://en.wikipedia.org/wiki/Heinz_von_Foerster}}  A historical and philosophical review of the work by von Foerster {\it et al.}\ is presented by Umpleby \cite{Umpleby-1990}.

This predicted ``doomsday'' seems to be actually arriving now.  An article \cite{WSJ-2024} ``Suddenly there aren’t enough babies: The whole world is alarmed'' by Greg Ip and Janet Adamy, published in The Wall Street Journal in May 2024, opens with a rather dramatic statement:
\begin{quote}
``The world is at a startling demographic milestone. Sometime soon, the global fertility rate will drop below the point needed to keep population constant. It may have already happened.''
\end{quote}
The article implies that global human population is about to start decreasing.

The question thus arises: what will happen in 2026?  Will human population go to infinity or begin to collapse?  In this paper, we re-examine human population dynamics using the currently available data from 10,000 BCE to 2023 CE.  We present quantitative fits of these empirical data using simple mathematical functions that we describe along the way together with the relevant literature.  The aim of these fits is to capture the dominant features of the data on a large scale, while avoiding fits of minor fluctuations and deviations on a small scale.  Highly detailed accuracy is not expected from these fits; yet they provide important insights into human population dynamics.  In Secs.~\ref{Sec:Exponential}--\ref{Sec:Peak}, we demonstrate that human population growth has gone through four stages described by distinct mathematical equations and culminating in the ever-increasing pace of growth analyzed in Sec.~\ref{Sec:Adiabatic}.  In Sec.~\ref{Sec:Fossil}, we compare historical data for human population and $\rm CO_2$ level in the atmosphere and demonstrate a linear relationship between them.  Moreover, we present a mathematical fit of the time-dependent level ${\rm CO_2}(t)$ and make a near-term forecast.  From a unified perspective, we introduce coupled differential equations for human population and $\rm CO_2$ emissions in Sec.~\ref{Sec:Dynamics} and further elaborate on them in Sec.~\ref{Sec:Complex} using a complex variable.  We compare our predictions with the reports by the Population Division of the United Nations (UN) in Sec.~\ref{Sec:Conclusion}.  Our paper leaves aside dynamics of economic and financial indicators, which were considered, e.g.,\ in Refs.~\cite{Johansen-2001,Sornette-2002,Safa-2016,Sibani-2020,Gleria-2024}, and spatial dynamics \cite{Kuretova-2010}.  More generally, the subject of our paper fits into a broader field of ``social physics'' \cite{Jusup-2022}, which applies mathematical methods developed in physics to social phenomena.

We conclude that, paradoxically, both claims cited at the beginning are inherently related and represent two sides of the same coin.  On the one hand, we confirm the hyperbolic growth pattern identified by von Foerster {\it et al.}~\cite{vonFoerster-1960}, but with a slightly different year $t_s=2030$ of the extrapolated singularity.  On the other hand, human population cannot become literally infinite.  Thus, the extrapolated singularity is avoided in close proximity and transforms into a peak in human population at $t_s=2030$.  After this peak, the world population is expected to decrease, which is consistent with the conclusions of Ref.~\cite{WSJ-2024}.  Moreover, our analysis also predicts a peak in the annual $\rm CO_2$ emissions at the same year $t_s=2030$ and a subsequent decline, while the overall atmospheric level of $\rm CO_2$ will continue to increase monotonically in time.

The primary source of the population data for our paper is the aggregator Our World in Data \cite{Our-World-in-Data}.  As indicated on their Web site, the historical estimates of population for the time period from 10,000 BCE to 1799 CE come from the History Database of the Global Environment (HYDE, v 3.3) of the Netherlands.  For the time period 1800--1949, the historical estimates are from Gapminder (v7).  For the time period 1950--2023, Our World in Data uses the population records from the UN World Population Prospects (2024 revision) and updates them biannually.  However, instead of the UN data, we use the population data from the World Bank (WB) available from 1960 to the latest year 2023 \cite{WorldBank}.  The WB population data are updated annually and represent the middle of the corresponding year.  Their single data variable SP.POP.TOTL for the total world population \cite{WorldBank} is easier to handle than a multitude of data entries contained in the UN files.   For some reason, the WB numbers for total population seem to be slightly lower in recent years than the UN numbers.  Our mathematical fitting applies to the WB data \cite{WorldBank}, but not necessarily to the UN data.

\section{Exponential growth from 10,000 to 3000 BCE}
\label{Sec:Exponential}

The growth rate $r=b-d$ of the population $N(t)$ in time $t$ is the difference between birth rate $b$ and death rate $d$.  Taking this rate to be constant,  Malthus\footnote{\url{https://en.wikipedia.org/wiki/Thomas_Robert_Malthus}} arrived at the linear equation
\begin{equation}  \label{N'=N} 
  \frac{dN}{dt} = \frac{N}{T},
\end{equation}
where we represent the rate $r=1/T$ in terms of the characteristic growth time $T$.  Equation (\ref{N'=N}) has the exponential solution
\begin{equation}  \label{N_exp} 
  N_{\rm exp}(t) = N_0\,e^{(t-t_0)/T},
\end{equation}
where $N_0$ is the population at an arbitrarily selected time $t_0$.  The exponential function (\ref{N_exp}) describes steady, unlimited growth, but it does not exhibit a singularity at any finite time.

\begin{figure}[t]
\centerline{\includegraphics[width=0.8\linewidth]{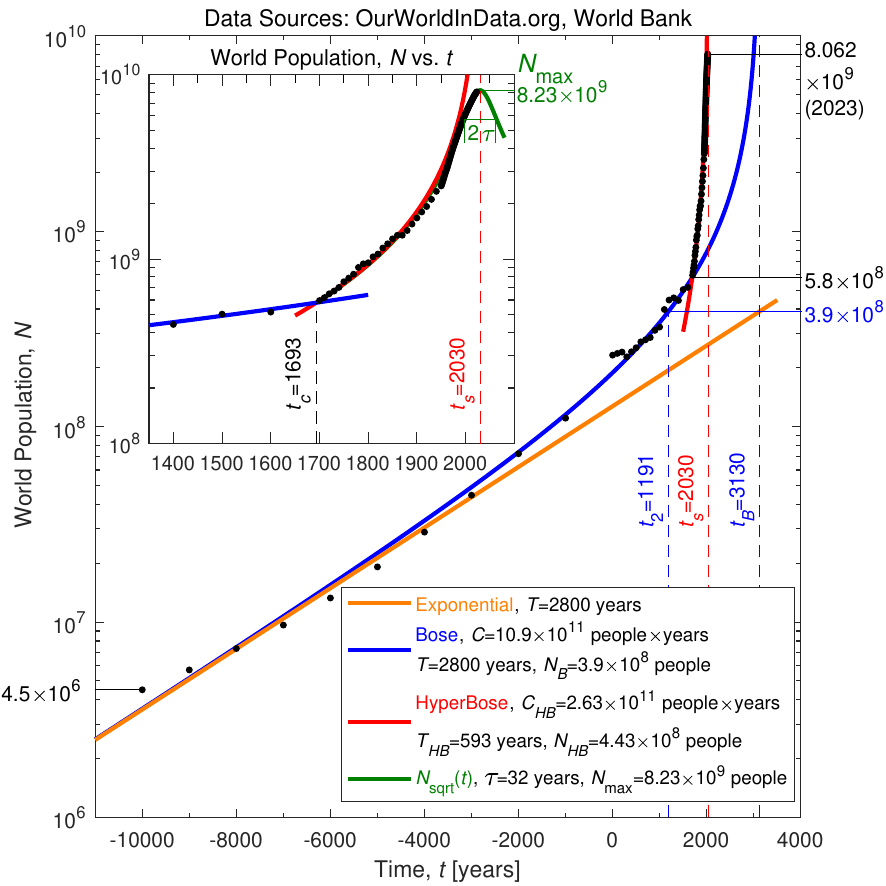}}
\caption{Black circles: the world population data $N(t)$.  Straight orange line: the exponential function (\ref{N_exp}).  Blue curves: the Bose function (\ref{N_Bose}).  Red curves: the HyperBose function (\ref{N_HyperBose}).  Green curve: the avoided-crossing function $N_{\rm sqrt}(t)$ (\ref{1/N_sqrt-HB}).  The singularity year is $t_s$ for the HyperBose fit and $t_B$ for the Bose fit, whereas $t_c$ marks the intersection of the two fits.  The Bose fit reaches its population scale $N_B$ in the year $t_2$.  Main message: Population initially grows exponentially (orange line), then super-exponentially (blue curves), and then switches to fast HyperBose growth (red curves) around $t_c=1693$.}
\label{Fig:Log}
\end{figure}

Figure~\ref{Fig:Log} shows human population $N(t)$ in the time range from 10,000 BCE (10k BCE) to 2023 CE from the data sources described in Sec.~\ref{Sec:Intro}.  The vertical scale for $N$ is logarithmic, and the data points are shown as black circles.  The endpoints indicate that the world population has increased from 4.5 million at 10k BCE to 8.062 billion at 2023 by more than three orders of magnitude.  

The main panel of Fig.~\ref{Fig:Log} demonstrates that initially, from 10k BCE, the data points approximately follow the straight orange line.  In the log-linear scale of the graph, the straight line indicates that human population growth was approximately exponential during the time period from 10k to 3k BCE, in agreement with the Malthusian Eq.~(\ref{N_exp}).  From the slope of the line, we extract the characteristic exponential growth time $T=2800$~years, the time in which the population increases by a factor of $e\approx2.7$.  This means that it took approximately three thousand years to roughly triple the population back then.  Had this exponential growth continued unchanged, the human population would be around 0.2 billion people today, which is roughly 40 times smaller than the actual number of about 8 billion, as seen in Fig.~\ref{Fig:Log}.

\section{Super-exponential Bose growth from 3000 BCE to 1700 CE}
\label{Sec:Bose}

Verhulst\footnote{\url{https://en.wikipedia.org/wiki/Pierre_François_Verhulst}} argued that exponential growth must slow down and eventually stop because of resource exhaustion.  Thus, he introduced the Verhulst equation with a negative quadratic term added to Eq.~(\ref{N'=N})
\begin{equation}  \label{N'=N-N2} 
  \frac{dN}{dt} = \frac{N}{T} - \frac{N^2}{S} 
  = \frac{N}{T} \left(1 - \frac{N}{N_{cc}} \right),
\end{equation}
where $N_{cc}=S/T$ is the carrying capacity.  The solution of Eq.~(\ref{N'=N-N2}) is called the logistic function\footnote{\url{https://en.wikipedia.org/wiki/Logistic_function}} 
\begin{equation}  \label{N_logistic} 
  N_{\rm logist}(t) = \frac{N_{cc}}{ e^{(t_l-t)/T} + 1 },
\end{equation}
where the time $t_l$ is selected so that $N(t_l)=N_{cc}/2$.  The logistic function (\ref{N_logistic}) starts from $N_{\rm logist}(t)\to0$ at $t\to-\infty$ and saturates at $N_{\rm logist}(t)\to N_{cc}$ at $t\to+\infty$, reaching the half-value $N_{cc}/2$ at $t=t_l$.

However, it is clear in the main panel of Fig.~\ref{Fig:Log} that the data points after 3k BCE systematically go \textit{above} the straight exponential line.  This indicates super-exponential growth, in complete disagreement with the sub-exponential growth prescribed by the Verhulst equation (\ref{N'=N-N2}) and the logistic function (\ref{N_logistic}).  To describe the accelerated growth, we change the sign of the quadratic term in Eq.~(\ref{N'=N-N2}) from negative to positive
\begin{equation}  \label{N'=N+N2} 
  \frac{dN}{dt} = \frac{N}{T} + \frac{N^2}{C} 
  = \frac{N}{T} \left(1 + \frac{N}{N_B} \right),
\end{equation}
where $C$ is a constant, and $N_B=C/T$.  For the lack of better terminology, we call Eq.~(\ref{N'=N+N2}) the ``anti-Verhulst equation.''  A derivation of Eq.~(\ref{N'=N+N2}) starting from two coupled differential equations is discussed in Sec.~\ref{Sec:Dynamics} along with relevant references.

The solution of Eq.~(\ref{N'=N+N2}) is given by the following function
\begin{equation}  \label{N_Bose} 
  N_{\rm Bose}(t) = \frac{N_B}{ e^{(t_B-t)/T} - 1 },
\end{equation}
which is mathematically analogous to the Bose-Einstein distribution function\footnote{\url{https://en.wikipedia.org/wiki/Bose-Einstein_statistics}} in statistical physics, if the variable $-t$ is replaced by energy $\varepsilon$, the parameter $-t_B$ by chemical potential $\mu$, and $T$ by temperature.  Thus, for the lack of better terminology, we call Eq.~(\ref{N_Bose}) the Bose function and use the subscript $B$ in Eqs.~(\ref{N'=N+N2}) and (\ref{N_Bose}), whereas the logistic function (\ref{N_logistic}) is similar to the Fermi-Dirac distribution.\footnote{\url{https://en.wikipedia.org/wiki/Fermi-Dirac_statistics}}

The Bose function (\ref{N_Bose}) monotonically increases starting from zero at $t\to-\infty$.  For a small $N\ll N_B$, the quadratic term $N^2$ can be neglected in Eq.~(\ref{N'=N+N2}), and so Eq.~(\ref{N_Bose}) reduces to the exponential function (\ref{N_exp})
\begin{equation}  \label{Bose->exp} 
  N_{\rm Bose}(t) \approx N_B\,e^{(t-t_B)/T} \quad {\rm for} \quad t_B-t\gg T.
\end{equation}
However, unlike the exponential function, the Bose function (\ref{N_Bose}) reaches a singularity $N_{\rm Bose}(t_B)\to\infty$ at the finite time $t_B$.  In the vicinity of this singularity, it reduces to the hyperbolic form
\begin{equation}  \label{Bose->hyp} 
  N_{\rm Bose}(t) \approx \frac{C}{t_B-t}  \quad {\rm for} \quad t_B-t\ll T.
\end{equation}
The hyperbolic regime (\ref{Bose->hyp}) applies when $N\gg N_B$, so that the term linear in $N$ in Eq.~(\ref{N'=N+N2}) can be neglected.

We find that the Bose function (\ref{N_Bose}) provides a good fit of super-exponential growth, as shown by the blue curve in the main panel of Fig.~\ref{Fig:Log}.  The Bose fit applies for the time range from 10k BCE to the year 1700 CE, after which population growth accelerates to a much faster pace, as discussed in Sec.~\ref{Sec:Hyperbolic}.  The Bose fit parameters are $T=2800$ years and $N_B=0.39$ billion people, and the singularity year is $t_B=3130$.  The straight orange line in Fig.~\ref{Fig:Log} represents the exponential limit (\ref{Bose->exp}) of the Bose fit (\ref{N_Bose}) for the early years.

Had this Bose growth continued unchanged, today's human population would be around 0.8 billion people, which is roughly 10 times smaller than the actual number of about 8 billion.  As indicated in the main panel of Fig.~\ref{Fig:Log}, the Bose fit reaches its characteristic population scale $N_{\rm Bose}(t_2)=N_B$ in the year $t_2=1191$, which, following from Eq.~(\ref{N_Bose}), is defined by the relation $e^{(t_B-t_2)/T}=2$.  The time $t_2$ separates the weakly super-exponential regime for $t<t_2$, where $N<N_B$ and the quadratic term in Eq.~(\ref{N'=N+N2}) is smaller than the linear term, from the strongly super-exponential regime for $t>t_2$ where $N>N_B$ and the quadratic term dominates.  It also follows from Eq.~(\ref{Bose->exp}) that the Bose function exceeds its exponential approximation at $t=t_2$ by a factor of two: $N_{\rm Bose}(t_2)=2N_{\rm exp}(t_2)$.  Moreover, $N_{\rm exp}(t_B)=N_B$ as indicated by the blue horizontal line in Fig.~\ref{Fig:Log}.

\section{Hyperbolic growth from 1700 to 2000}
\label{Sec:Hyperbolic}

In the main panel and the zoomed-in inset of Fig.~\ref{Fig:Log}, the black data points veer up sharply above the blue curve after the year 1700, thus indicating a switch to a much faster population growth.  We find that the population data $N(t)$ from 1700 to 2000 can be well fitted using the Bose function (\ref{N_Bose}) with different parameters $T_{H\!B}=593$ years, $N_{H\!B}=0.443$ billion people, and the singularity year $t_s=2030$:
\begin{equation}  \label{N_HyperBose} 
  N_{\rm HyperBose}(t) = \frac{N_{H\!B}}{ e^{(t_s-t)/T_{H\!B}} - 1 }.
\end{equation}
We call this regime HyperBose and use the label $H\!B$ to distinguish it from the slower Bose growth before 1700.  The HyperBose fit (\ref{N_HyperBose}) is shown by the red curves in Fig.~\ref{Fig:Log}.  The inset in Fig.~\ref{Fig:Log} indicates that the slow (blue) and the fast (red) Bose fits intersect at the crossover year $t_c=1693$, marking a change of regime.

Soon after 1700, the HyperBose fit (\ref{N_HyperBose}) becomes hyperbolic, as proposed by von Foerster {\it et al.}\ \cite{vonFoerster-1960},
\begin{equation}  \label{N_hyp} 
  N_{\rm hyper}(t) = \frac{C_{H\!B}}{t_s-t}, \qquad {\rm where} 
  \qquad C_{H\!B}=T_{H\!B}N_{H\!B}. 
\end{equation}
Equation (\ref{N_hyp}) is a solution of the nonlinear differential equation
\begin{equation}  \label{N'=N2} 
  \frac{dN}{dt} = \frac{N^2}{C_{H\!B}},
\end{equation}
which can be equivalently rewritten for the inverse population $1/N$
\begin{equation}  \label{1/N'} 
  \frac{d}{dt} \frac{1}{N} = - \frac{1}{C_{H\!B}}.
\end{equation}
Equation (\ref{1/N'}) has an obvious mathematical solution, which is the inverse of Eq.~(\ref{N_hyp}), 
\begin{equation}  \label{1/N_hyp} 
  \frac{1}{N_{\rm hyper}(t)} = \frac{t_s-t}{C_{H\!B}}.
\end{equation}
Equation (\ref{1/N_hyp}) shows that the inverse population $1/N$ decreases linearly in time $t$, until it crosses zero $1/N_{\rm hyper}(t_s)\to0$ at $t=t_s$, indicating a singularity $N_{\rm hyper}(t_s)\to\infty$.  Past the singularity, $N_{\rm hyper}(t)$ formally becomes negative for $t>t_s$.

\begin{figure}[t]
\centerline{\includegraphics[width=0.8\linewidth]{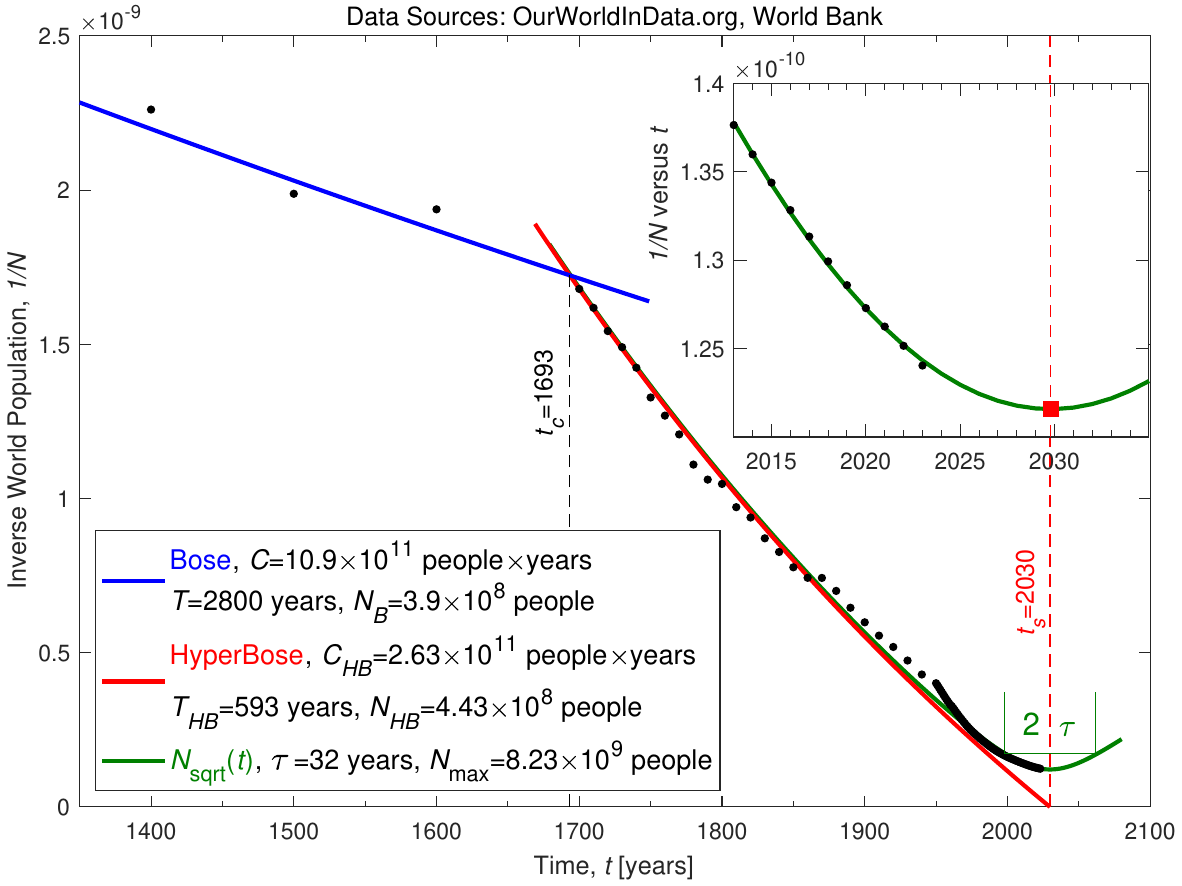}}
\caption{The inverse world population $1/N$ versus time $t$.  Notation is the same as in Fig.~\ref{Fig:Log}.  The red HyperBose curve (\ref{N_HyperBose}) extrapolates linearly to zero, indicating a projected singularity at $t_s=2030$.  But the green curve for $1/N_{\rm sqrt}(t)$ (\ref{1/N_sqrt-HB}) fits the data points avoiding zero crossing and reaches minimum at $t_s$, thus transforming a singularity into a population peak.}
\label{Fig:1/N}
\end{figure}

To illustrate hyperbolic behavior, we plot the inverse population $1/N$ (black dots) versus time $t$ in Fig.~\ref{Fig:1/N}.  The red curve in the main panel shows the HyperBose fit (\ref{N_HyperBose}) for $1/N$, which becomes an approximately straight line after 1800, in agreement with Eq.~(\ref{1/N_hyp}).  The intersection point of the red line with the time axis yields the hyperbolic singularity year $t_s=2030$, which is close to the predicted year 2026 by von Foerster {\it et al.}\ \cite{vonFoerster-1960}.  The inverse slope of the straight red line gives the parameter $C_{H\!B}=2.63\times10^{11}$ people$\,\times\,$years, which is close to the value $1.8\times10^{11}$ found in Ref.~\cite{vonFoerster-1960}.  Thus, the current data on human population essentially confirm the hyperbolic growth pattern with the year $t_s=2030$ of extrapolated singularity, as envisioned by von Foerster {\it et al.}\ \cite{vonFoerster-1960} roughly 65 years ago.

However, it is important to emphasize that the red HyperBose fit in Fig.~\ref{Fig:1/N} is applicable only after the year 1700, not during all of human history, as was assumed by von Foerster and other authors.  The main panel of Figure~\ref{Fig:Log} shows that the red HyperBose fit was preceded by the blue Bose fit for more than 11 thousand years of human history.  The black horizontal line in the main panel of Fig.~\ref{Fig:Log} shows that the human population at the crossover year was $N(t_c)=0.58$ billion, which is about 1.5 times greater than $N_B=0.39$ billion.  Thus, the slow Bose function was already in the hyperbolic limit (\ref{Bose->hyp}) at $t_c$ and appears as the straight blue line in the graph of $1/N(t)$ in Fig.~\ref{Fig:1/N}.  But the slope of the blue line is about four times smaller than the slope of the red hyperbolic line, because the Bose parameter $C=TN_B=10.9\times10^{11}$ people$\,\times\,$years is about four times greater than the HyperBose parameter $C_{H\!B}=2.63\times10^{11}$ people$\,\times\,$years.  Thus, the speed $dN/dt$ of human population growth quadrupled near the crossover year $t_c=1693$ upon switching to the HyperBose regime.   We attribute this switch to the onset of the Industrial Revolution and the transition to massive use of fossil fuels, as discussed in more detail in Sec.~\ref{Sec:Fossil}.

Figure~\ref{Fig:1/N} demonstrates a practical method for visualization of a diverging variable $N(t)$ by making a graph of its inverse $1/N(t)$.  We are not aware of any paper in the demographic literature presenting a graph of $1/N$ versus $t$, which would reveal hyperbolic growth.  But this method is commonly used in physics, particularly for phase transitions.  As an example, Figure 1 in Ref.~\cite{Monceau-2001} shows experimental data for the inverse of the dielectric permittivity $\epsilon$ versus temperature $T$.  While $\epsilon\propto 1/|T-T_0|$ goes to infinity at the temperature $T_0$ of a ferroelectric phase transition, its inverse $1/\epsilon\propto|T-T_0|$ goes to zero linearly, which is clearly demonstrated in Figure 1 of Ref.~\cite{Monceau-2001}.  

Alternatively, Refs.~\cite{vonFoerster-1960} and \cite{Johansen-2001} visualized human population data using log-log coordinates of $\log N$ versus $\log(t_s-t)$ in an attempt to find a power-law divergence $N(t)\propto1/(t_s-t)^\alpha$.  While this method is suitable for determining the exponent $\alpha$, identifying the singularity time $t_s$ is problematic.  Nevertheless, by analyzing the population data available in 1960, von Foerster {\it et al.}\ \cite{vonFoerster-1960} concluded that the best fit is obtained for $\alpha=1$, i.e.,\ by the simple hyperbolic function (\ref{N_hyp}), with $t_s=2026$.  Subsequently, Johansen and Sornette \cite{Johansen-2001} analyzed population data available  up to 2000.  They considered three arbitrarily selected singularity years $t_s=2030$, 2040, and 2050 and found the corresponding exponents $\alpha=0.88$, 1.01, and 1.14, as shown in their Table 1.  The middle exponent is very close to 1, whereas the other two deviate from 1 by only -12\% and +14\%.  These small deviations of $\alpha$ from 1 are consistent with a simple hyperbolic law within the accuracy of the fits and scattering of data points, and are insufficient to claim a generic power law.  Moreover, by including the data points after 2000 (not available to Johansen and Sornette) and regularizing the singularly as discussed in Sec.~\ref{Sec:Peak} (not attempted by Johansen and Sornette), we find that the best fitting gives $\alpha=1$ and $t_s=2030$ in Eq.~(\ref{N_hyp}), as illustrated by the red line in Fig.~\ref{Fig:1/N}.

\section{Slow-down after 2000 and a population peak in 2030}
\label{Sec:Peak}

Human population cannot become literally infinite.  It is therefore necessary to introduce some kind of cutoff near the singularity in the hyperbolic formula (\ref{N_hyp}).  Indeed, in the main panel of Fig.~\ref{Fig:1/N}, the black data points for $1/N(t)$ after the year 2000 veer up above the straight red line of the hyperbolic fit, thus avoiding zero-crossing and the singularity.

Mathematically, we need to prevent the inverse population $1/N$ in Eq.~(\ref{1/N_hyp}) from crossing zero and becoming negative, as this would be unphysical. A mathematical procedure for avoided crossing\footnote{\url{https://en.wikipedia.org/wiki/Avoided_crossing}} is well known in the context of quantum physics.  Thus, we propose the following regularization of Eq.~(\ref{1/N_hyp})
\begin{equation}  \label{1/N_sqrt-L} 
  \frac{1}{N_{\rm sqrtL}(t)} = \frac{\sqrt{(t_s-t)^2+\tau^2}}{C_{H\!B}},
\end{equation}
where $\tau$ is a cutoff time.  Equation (\ref{1/N_sqrt-L}) reduces to the hyperbolic formula (\ref{1/N_hyp}) for $t_s-t\gg\tau$, but it has a nonzero minimal value at $t=t_s$ and remains positive at all times.  Correspondingly, the population is given by the square root of the Lorentzian function, hence the label sqrtL,
\begin{equation}  \label{N_sqrt-L} 
  N_{\rm sqrtL}(t) = \frac{C_{H\!B}}{\sqrt{(t_s-t)^2+\tau^2}}.
\end{equation}
An alternative derivation of Eq.~(\ref{N_sqrt-L}) is presented in Sec.~\ref{Sec:Complex}.  According to Eq.~(\ref{N_sqrt-L}), human population is expected to reach a peak at $t=t_s$ with the maximal value
\begin{equation}  \label{N_max} 
  N_{\rm max} = \frac{C_{H\!B}}{\tau}
\end{equation}
and to decrease hyperbolically for $|t-t_s|\gg\tau$.  At $t=t_s\mp\tau$, the population level is $N_{\rm max}/\sqrt{2}$ in Eq.~(\ref{N_sqrt-L}), thus the parameter $2\tau$ gives the time width of the peak.

More generally, we use the following regularization of the singularity in the HyperBose function (\ref{N_HyperBose})
\begin{equation}  \label{1/N_sqrt-HB} 
  \frac{1}{N_{\rm sqrt}(t)} = 
  \sqrt{ \frac{1}{N^2_{\rm HyperBose}(t)} + \frac{1}{N^2_{\rm max}} },
\end{equation}
where $N_{\rm max}$ is treated as a fitting parameter.  The regularized function $N_{\rm sqrt}(t)$ smoothly interpolates between $N_{\rm HyperBose}(t)$ for $|t-t_s|\gg\tau$ and $N_{\rm sqrtL}(t)$ for $|t-t_s|\ll T_{H\!B}$, where
\begin{equation}  \label{tau} 
  \tau = T_{H\!B} \, \frac{N_{H\!B}}{N_{\rm max}}.
\end{equation}

The green curve in the main panel of Fig.~\ref{Fig:1/N} demonstrates a good fit of the black data points for $1/N(t)$ using the function $1/N_{\rm sqrt}(t)$ from Eq.~(\ref{1/N_sqrt-HB}) with the fitting parameters $\tau=32$~years and $N_{\rm max}=8.23$~billion people.  The green curve for $1/N_{\rm sqrt}(t)$ reduces to the red curve for $1/N_{\rm HyperBose}(t)$ before 2000, but also fits the data points veering up above the red line after 2000.  The inset in Fig.~\ref{Fig:1/N} zooms into recent years and shows that the green curve for $1/N_{\rm sqrt}(t)$ has a typical parabolic shape near its predicted minimum at $t_s=2030$.  (Similarly in physics, the graphs of $1/\epsilon$ versus $T$ in Figure 1 of Ref.~\cite{Monceau-2001} also show avoided crossing for several materials, quite reminiscent of the green curve in the main panel of our Fig.~\ref{Fig:1/N}.)  

Mathematically, since $1/N_{\rm sqrt}(t)>1/N_{\rm HyperBose}(t)$ in Eq.~(\ref{1/N_sqrt-HB}), the green curve in the main panel of Fig.~\ref{Fig:1/N} can fit the black data points only if the red line for its hyperbolic asymptote (\ref{1/N_hyp}) passes slightly below the data points.  This requirement pushes the singularity time $t_s$ to the relatively early year 2030 in our fit.  In contrast, the later years $t_s=2040$ and $t_s=2050$, also considered by Johansen and Sornette \cite{Johansen-2001}, allow for decent fits of the data points up to 2000, but subsequent data points after 2000 cannot be fitted using the square-root regularization (\ref{1/N_sqrt-L}) for any choice of $\tau$.

\begin{figure}[t]
\centerline{\includegraphics[width=0.8\linewidth]{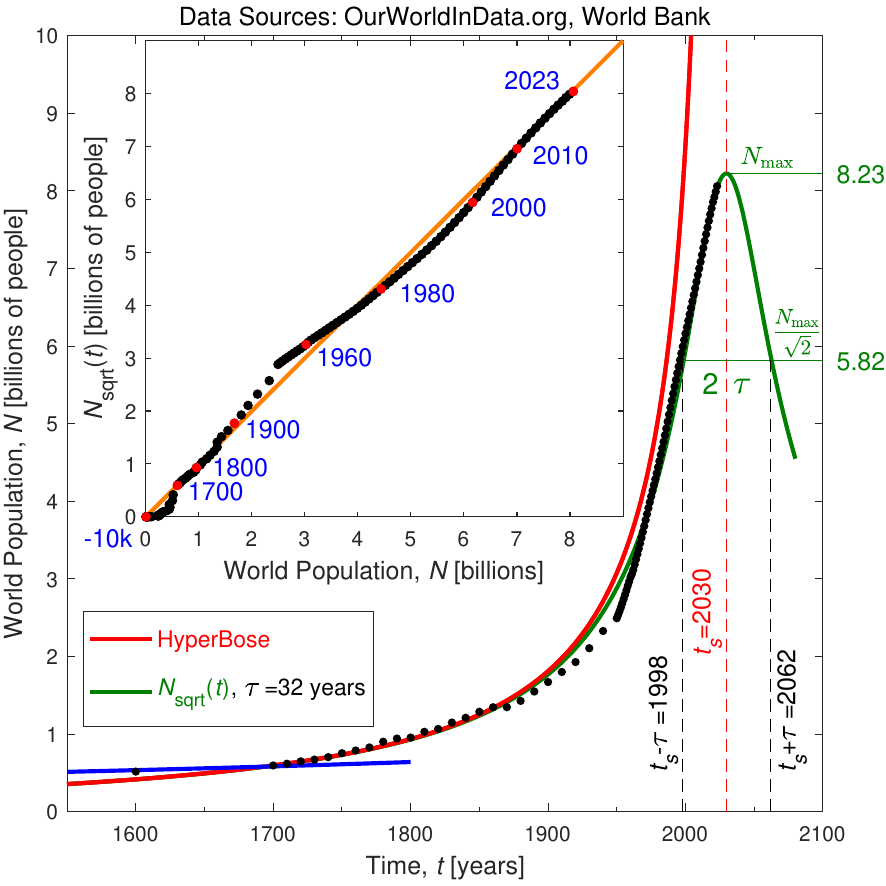}}
\caption{Main panel: The world population data points $N(t)$.  The red HyperBose curve (\ref{N_HyperBose}) goes to infinity at $t_s=2030$, but the green avoided-crossing fit $N_{\rm sqrt}(t)$ (\ref{1/N_sqrt-HB}) has a peak at $t_s$.  The predicted population maximum is $N_{\rm max}$, whereas the time width of the peak is $2\tau=64$ years at the height $N_{\rm max}/\sqrt{2}$.  Inset: A parametric plot of the population data $N(t)$ versus the theoretical fit $N_{\rm sqrt}(t)$.  Collapse of the data points on the straight diagonal orange line indicates a good fit from 1700 to 2023. }
\label{Fig:N-peak}
\end{figure}

The green curve for $N_{\rm sqrt}(t)$ in the main panel of Fig.~\ref{Fig:N-peak} shows in linear scale how the singularity in $N_{\rm HyperBose}(t)$ (red curve) is regularized and transformed into a peak in human population at $t_s=2030$.  The maximum population is projected to be $N_{\rm max}=8.23$ billion people, with a decrease to $N_{\rm max}/\sqrt{2}=5.82$ billion at $t_s-\tau=1998$ and $t_s+\tau=2062$ already by the middle of the 21st century.  The inset in Fig.~\ref{Fig:N-peak} compares the empirical data for human population $N(t)$ and the proposed theoretical fit $N_{\rm sqrt}(t)$ in a parametric plot, where time $t$ serves a parameter.  For each value of $t$ (some marked in blue), a data point on the inset has the horizontal coordinate $N(t)$ and the vertical coordinate $N_{\rm sqrt}(t)$.  The inset shows that the data points fall close to the straight diagonal orange line from 1700 to 2023, for more than three centuries.  This observation indicates that $N(t)\approx N_{\rm sqrt}(t)$ and thus confirms a good agreement between our fit and the data.  This method of comparison between empirical data and a theoretical hypothesis is similar in spirit to the Q-Q\footnote{\url{https://en.wikipedia.org/wiki/Q-Q_plot}} and P-P\footnote{\url{https://en.wikipedia.org/wiki/P-P_plot}} plots used for probability distributions. 

Of course, one may consider other ways of mathematically regularizing hyperbolic growth.  Kapitza \cite{Kapitza-1996} proposed that the \textit{derivative} of the hyperbolic function (\ref{N_hyp}) should be modified as follows
\begin{equation}  \label{Kapitza'} 
  \frac{dN_{\rm Kapitza}}{dt} = \frac{C}{(t_s-t)^2+\tau^2}.
\end{equation}
An integral of Eq.~(\ref{Kapitza'}) gives the following dependence of population on time
\begin{equation}  \label{Kapitza} 
  N_{\rm Kapitza}(t) = \frac{C}{\tau}{\rm arccot}\left(\frac{t_s-t}{\tau}\right),
\end{equation}
where ${\rm arccot}(x)$ is the inverse function\footnote{\url{https://en.wikipedia.org/wiki/Inverse_trigonometric_functions}} to $\cot(x)=\cos x/\sin x$.  Kapitza's function is somewhat similar to the logistic function (\ref{N_logistic}) in that it starts from zero at $t\to-\infty$ and saturates at a constant $N_{\rm Kapitza}(t)\to C\pi/\tau$ when $t\to+\infty$.  However, the two functions then differ when $t_s-t\gg\tau$: while the logistic function assumes exponential growth for this regime, Kapitza's function becomes hyperbolic.

Despite its ability to account for hyperbolic growth, Kapitza's proposed function (\ref{Kapitza}) does not agree with empirical data.  This is because, in Eq.~(\ref{Kapitza'}), the time derivative of population $dN/dt$ -- not the population $N$ itself -- should peak at the hyperbolic singularity time $t_s$.  In Ref.~\cite{Kapitza-1996} published in 1996,  Kapitza predicted a peak in the speed $dN/dt$ of population growth around the year $2007$.  He later revised this prediction to the year 2000, after $dN/dt$ had already peaked and started to decrease, as can be seen in the UN data shown in Figure 4 of his last paper \cite{Kapitza-2013} of 2013.  Regardless, whether 2000 or 2007, neither one of Kapitza's estimates for $t_s$ is compatible with the \textit{future} value $t_s=2030$ obtained from the red straight-line fit of $1/N(t)$ in the main panel of our Fig.~\ref{Fig:1/N}.  Kapitza did not realize this contradiction, because he never produced a graph of inverse population $1/N$ versus time $t$.  We conclude that empirical data do not agree with Kapitza's proposal (\ref{Kapitza}) for monotonic increase in human population, but are compatible with our Eq.~(\ref{1/N_sqrt-HB}), which predicts a peak and a subsequent decrease in population, as shown in Fig.~\ref{Fig:N-peak}.

\section{From the adiabatic to the diabatic regime}
\label{Sec:Adiabatic}

In this Section, we discuss the rate $r$ of population growth, defined by the logarithmic derivative 
\begin{equation}  \label{rate} 
  r = \frac{1}{N}\frac{dN}{dt} = \frac{d\ln N}{dt},
\end{equation}
and the corresponding growth time $1/r$. 

As described in Sec.~\ref{Sec:Exponential}, some ten thousand years ago, $N(t)$ was exponential with an approximately constant growth time $T=2800$~years. This growth time greatly exceeded the average life span of a human being, so any societal changes experienced by an individual during his or her lifetime tended to be very gradual.  Thus, during this period of exponential growth in population, human beings found themselves in the adiabatic (quasi-static) regime,\footnote{\url{https://en.wikipedia.org/wiki/Adiabatic_theorem\#Diabatic_vs._adiabatic_processes}} where individuals hardly noticed systematic changes during their lifetimes.  This observation aligns with the fact that an exponentially growing function is an eigenfunction of the time-translation operator, and thus can be thought of as representing a time-translation symmetry of human society, which effectively prevents any particular moment in time from being unique over another.  

However, the situation is very different in the hyperbolic regime described by Eq.~(\ref{N_hyp}), where a time-translation symmetry is broken by the singularity time $t_s$.  In stark contrast to the constant rate for exponential growth, the hyperbolic growth rate diverges at $t\to t_s$:
\begin{equation}  \label{rate_hyp} 
  r_{\rm hyper} = \frac{1}{t_s-t}.
\end{equation}
Correspondingly, the characteristic growth time decreases linearly in time and extrapolates to zero at $t\to t_s$:
\begin{equation}  \label{1/rate_hyp} 
  \frac{1}{r_{\rm hyper}} = t_s-t.
\end{equation}
Equation (\ref{1/rate_hyp}) represents a scale-free process, where three hundred years ago, the growth time was 300 years; two hundred years ago was 200 years, etc.  At present time, this mathematical trend indicates a growth time shorter than the average human life span.  Thus, we conclude that human population growth has evolved from the adiabatic regime some 10k years ago to the diabatic (or anti-adiabatic) regime now, where human society changes on a time scale shorter than an individual's lifetime. 

However, it is well known for various self-similar processes in physics, e.g., for turbulence,\footnote{\url{https://en.wikipedia.org/wiki/Turbulence\#Kolmogorov's_theory_of_1941}} that macroscopic scaling breaks down at a microscopic scale.  This means that, while scaling may be applicable over a wide range (the inertial range for turbulence), it breaks down when an intrinsic microscopic scale of the problem is reached (for turbulence, this is the dissipative viscous scale).  From this perspective, one can imagine that a microscopic cutoff scale for the macroscopic growth time (\ref{1/rate_hyp}) of global population is the life span of an individual human.  When the time distance $t_s-t$ to the extrapolated singularity becomes comparable with the life span of a human, the global growth process is expected  to deviate from its hyperbolic behavior in order to avoid a literal mathematical singularity.  As we argue in Sec.~\ref{Sec:Peak}, this regularization is described by the avoided-crossing formula (\ref{N_sqrt-L}), which gives the growth rate
\begin{equation}  \label{rate_sqrt} 
  r_{\rm sqrtL} = \frac{t_s-t}{(t_s-t)^2+\tau^2}
\end{equation}
and the growth time
\begin{equation}  \label{1/rate_sqrt} 
  \frac{1}{r_{\rm sqrtL}} = t_s-t + \frac{\tau^2}{t_s-t}.
\end{equation}
The magnitude of the growth time in Eq.~(\ref{1/rate_sqrt}) reaches a nonzero minimal value min$(1/|r_{\rm sqrtL}|)=2\tau$ at $t=t_s\mp\tau$.  The value of $\tau=32$~years found in Sec.~\ref{Sec:Peak} gives $2\tau=64$~years, which is, indeed, comparable with the life span of a human.  Thus, over the course of human history, the characteristic time of population growth has decreased by roughly two orders of magnitude, from $T=2800$~years to $2\tau=64$~years.

\section{Population growth and extraction of fossil fuels}
\label{Sec:Fossil}

\begin{figure}[t]
\centerline{\includegraphics[width=0.8\linewidth]{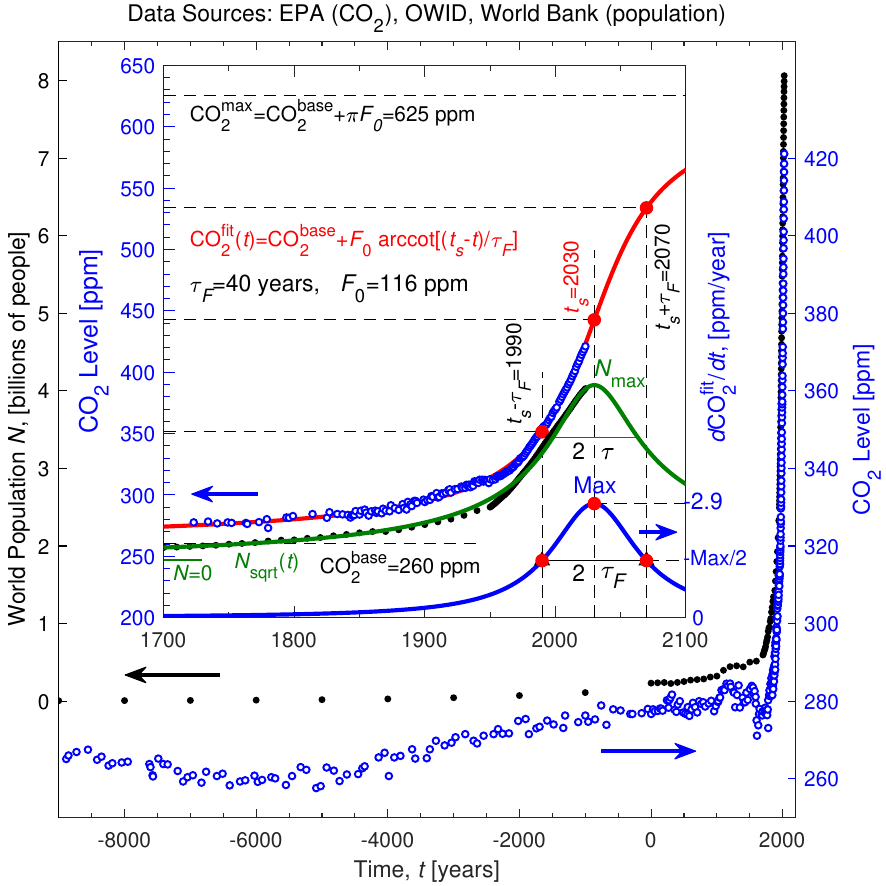}}
\caption{Main panel: The atmospheric carbon dioxide concentration ${\rm CO_2}(t)$ (open blue circles) and the human population $N(t)$ (solid black circles) both start growing rapidly at the onset of the Industrial Revolution.  Inset: The same data points exhibit parallel growth from 1700.  The green curve for $N_{\rm sqrt}(t)$ (\ref{1/N_sqrt-HB}) and the red curve for ${\rm CO_2^{fit}}(t)$ (\ref{CO2(t)}) are theoretical fits for population and carbon dioxide level.  The blue curve at the bottom shows the time derivative $d{\rm CO_2^{fit}}/dt$ (\ref{dCO2/dt}).  It represents a Lorentzian peak in the annual $\rm CO_2$ emissions at $t_s=2030$ with the time width of $\tau_F=40$ years.}
\label{Fig:CO2-N}
\end{figure}

As discussed in Sec.~\ref{Sec:Hyperbolic}, the fast HyperBose growth started around 1700, which roughly coincides with the onset of the Industrial Revolution and the transition to massive use of fossil fuels.  At a qualitative level, a correlation between extraction of fossil fuels and human population growth has been recognized in the literature, e.g.,\ in Refs.~\cite{Safa-2016,Husler-2014,Bongaarts-2018}.  Here we present a quantitative comparison between human population $N(t)$ and carbon dioxide concentration ${\rm CO_2}(t)$ in the atmosphere, measured in ppm (parts per million).  We use the ${\rm CO_2}(t)$ data from Antarctic ice cores up to 1958 and from direct measurements at Mauna Loa starting 1959, as compiled by the Environmental Protection Agency (EPA) \cite{EPA-CO2} from underlying datasets.

In the main panel of Fig.~\ref{Fig:CO2-N}, the open blue circles show ${\rm CO_2}(t)$ from 9k BCE to 2023 CE.  At the right edge of the panel, we observe a sharp increase in the last 300 years, known as the ``hockey stick graph.''\footnote{\url{https://en.wikipedia.org/wiki/Hockey_stick_graph_(global_temperature)}}  In the same main panel, the solid black circles show human population $N(t)$, which was close to zero on a linear scale for more than ten thousand years and then sharply increased in the last 300 years, thus exhibiting a similar ``hockey stick graph.''  We see that human population and atmospheric $\rm CO_2$ level started growing explosively at about the same time.  Moreover, the open blue and solid black data points in the zoomed-in inset of Fig.~\ref{Fig:CO2-N} confirm that ${\rm CO_2}(t)$ and $N(t)$ increase in a very similar manner after 1700.  

\begin{figure}[t]
\centerline{\includegraphics[width=0.8\linewidth]{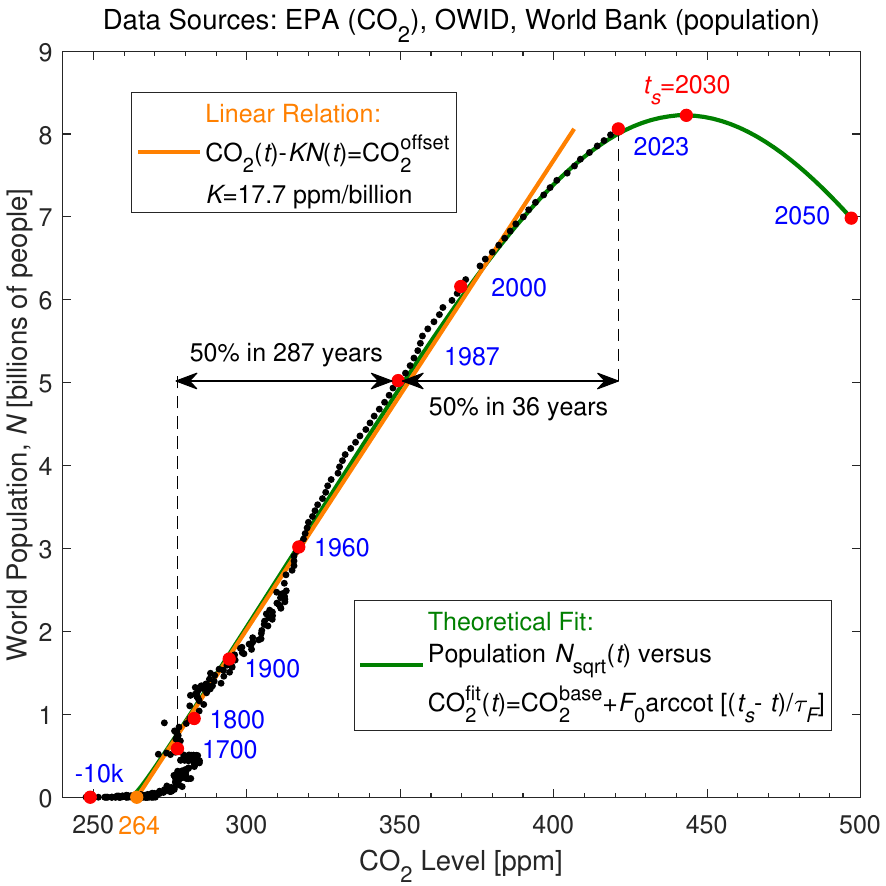}}
\caption{Black circles: A parametric plot of the empirical data for ${\rm CO_2}(t)$ versus human population $N(t)$.  Green curve: a theoretical plot of ${\rm CO_2^{fit}}(t)$ (\ref{CO2(t)}) versus $N_{\rm sqrt}(t)$ (\ref{1/N_sqrt-HB}), showing a good agreement with the data.  The straight orange line indicates a linear relation between population and carbon dioxide level from 1700 to 2000.  Of the total ${\rm CO_2}$ increase from 1700 to 2023, the second half (50\%) was emitted after 1987 in only 36 years, whereas it took 287 years to emit the first half.}
\label{Fig:QQ}
\end{figure}

\begin{figure}[t]
\centerline{\includegraphics[width=0.8\linewidth]{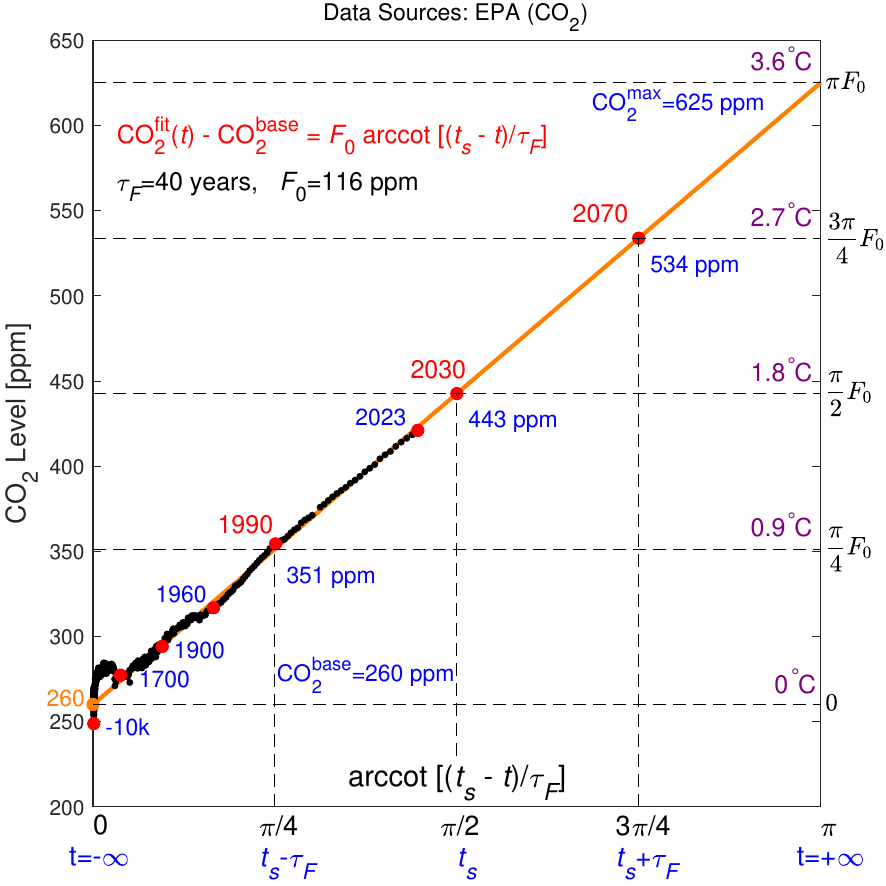}}
\caption{Parametric plot of the data points for the carbon dioxide concentration ${\rm CO_2}(t)$ versus the inverse trigonometric function ${\rm arccot}[(t_s-t)/\tau_F]$.  The straight orange line indicates goodness of the theoretical fit ${\rm CO_2^{fit}}(t)$ (\ref{CO2(t)}) from 1700 to the present time 2023.  The $\rm CO_2$ levels at the times $t=-\infty$, $t_s-\tau_F$, $t_s$, $t_s+\tau_F$, and $+\infty$ are spaced equidistantly.  The right vertical axis shows the corresponding increments in global temperature using the conversion factor of $100\:{\rm ppm}\approx 1^\circ\:{\rm C}$.}
\label{Fig:CO2-arccot}
\end{figure}

For quantitative comparison, Fig.~\ref{Fig:QQ} presents a parametric plot, where time $t$ serves a parameter.  For each value of $t$ (some marked in blue), a data point has the horizontal coordinate ${\rm CO_2}(t)$ and the vertical coordinate $N(t)$.  When $t$ runs from 10k BCE to 2023 CE, the data points yield a parametric plot of human population versus atmospheric carbon dioxide concentration.  Because the data for ${\rm CO_2}(t)$ are available at a higher time frequency than for population $N(t)$, we interpolate the sparser data for $N(t)$ to match the same time frequency.  Thus, the interpolated data points for $N(t)$ shown in Figs.~\ref{Fig:QQ} and \ref{Fig:CO2-arccot} have higher density than the raw data for population shown in Figs.~\ref{Fig:Log}--\ref{Fig:CO2-N}. 

Figure~\ref{Fig:QQ} demonstrates that the data points approximately follow the straight orange line for the three centuries from 1700 to 2000.  This straight line indicates a linear relation between ${\rm CO_2}(t)$ and $N(t)$:
\begin{equation}  \label{CO2-HN} 
  {\rm CO_2}(t) - K N(t) = {\rm CO_2^{offset}}.
\end{equation}
The parameters $K$ and $\rm CO_2^{offset}$ are given in Eqs.~(\ref{K}) and (\ref{const}) at the end of this Section.  Equation~(\ref{CO2-HN}) and the straight line in Fig.~\ref{Fig:QQ} demonstrate that the sharp increase in carbon dioxide concentration after 1700 is directly proportional to the increase in human population and, thus, is undoubtedly attributed to humans' massive use of fossil fuels following the Industrial Revolution. The linear correlation (\ref{CO2-HN}) is a quantitative counterpart to the qualitative observations made in the literature \cite{Safa-2016,Husler-2014,Bongaarts-2018}.  A mathematical consequence of the linear relation (\ref{CO2-HN}) is that the carbon dioxide level ${\rm CO_2}(t)$ grows hyperbolically in time (relative to an offset level), because the population $N(t)$ exhibits such a hyperbolic growth.  While the latter was uncovered in Ref.~\cite{vonFoerster-1960}, the evidence for hyperbolic growth also in ${\rm CO_2}(t)$ presented by Eq.~(\ref{CO2-HN}) was not realized before, to the best of our knowledge.

However, the linear relation (\ref{CO2-HN}) applies only after 1700, but not before, as seen in  Fig.~\ref{Fig:QQ}.  Between 10k BCE and 1700 CE, human population was rather small, whereas $\rm CO_2$ varied naturally between roughly 250 and 285 ppm, clearly unrelated to population.  Figure~\ref{Fig:QQ} also indicates that, of the total $\rm CO_2$ increase since the Industrial Revolution from 1700 to 2023, the second half (50\%) was emitted after 1987 in only 36 years, whereas it had taken 287 years to emit the first half (50\%) from 1700 to 1987.\footnote{Incidentally, the year of my Ph.D.\ is 1987, so a half of the total $\rm CO_2$ emissions happened during my academic career.}

Looking at Fig.~\ref{Fig:QQ}, we also see that the data points recently deviate downward from the straight orange line, thus indicating a breakdown of the linear relation (\ref{CO2-HN}) after 2000.  This downward deflection can be understood by examining the data points for $N(t)$ and ${\rm CO_2}(t)$ after 2000 in the inset of Fig.~\ref{Fig:CO2-N}.  The solid black circles and the green curve for $N_{\rm sqrt}(t)$ show that human population growth decelerates after 2000, when $N(t)$ is approaching a peak in 2030.  In contrast, the open blue circles show that  the carbon dioxide level ${\rm CO_2}(t)$ continues to increase unabated.  This is because the emitted $\rm CO_2$ stays in the atmosphere for a very long time, between 300 and 1000 years \cite{CO2-stay}.  Thus, the carbon dioxide level ${\rm CO_2}(t)$ cannot possibly decrease on a shorter time scale and can only increase monotonically without a peak.  As a result, the separation between the curves for ${\rm CO_2}(t)$ and $N(t)$ in the inset of Fig.~\ref{Fig:CO2-N} increases after 2000.  Consequently, the parametric plot of $N$ versus $\rm CO_2$ in Fig.~\ref{Fig:QQ} deflects downward after 2000.

In Sec.~\ref{Sec:Peak} we demonstrated that the mathematical function $N_{\rm sqrt}(t)$ in Eq.~(\ref{1/N_sqrt-HB}) gives a good fit of the empirical data $N(t)$ for human population since 1700, as validated by the inset in Fig.~\ref{Fig:N-peak}.  A question arises whether a similar mathematical fit can be obtained for the carbon dioxide level ${\rm CO_2}(t)$ since 1700.  A mathematical fitting function ${\rm CO_2^{fit}}(t)$ must satisfy the following conditions.  First, it must be a monotonically increasing function of time without a peak, as discussed above.  Second, the function must exhibit a hyperbolic behavior in the past for $t_s-t\gg\tau$, because of the linear relation (\ref{CO2-HN}) with $N(t)$ growing hyperbolically.  The third condition is not strictly necessary, but can be taken as an optimistic assumption that the carbon dioxide level eventually saturates at a constant level ${\rm CO_2^{fit}}(t)\to\rm CO_2^{max}$ in distant future at $t-t_s\gg\tau$.  A mathematical function satisfying the all three conditions is the arccot function (\ref{Kapitza}) proposed by Kapitza \cite{Kapitza-1996} for population $N(t)$.  As explained at the end of Sec.~\ref{Sec:Peak}, Kapitza's proposal does not work for population, but it can be a good candidate for ${\rm CO_2^{fit}}(t)$.  Thus we propose the following fit for the carbon dioxide level
\begin{equation}  \label{CO2(t)} 
  {\rm CO_2^{fit}}(t) = {\rm CO}_2^{\rm base} 
  + F_0 \, {\rm arccot}\left(\frac{t_s-t}{\tau_F}\right).
\end{equation}
Here the time width $\tau_F$ and the coefficients ${\rm CO}_2^{\rm base}$ and $F_0$ are fitting parameters.  But the time $t_s=2030$ of the infection point in Eq.~(\ref{CO2(t)}) is taken to be the same as $t_s$ for $N_{\rm HyperBose}(t)$ in Eq.~(\ref{N_HyperBose}) in order to match hyperbolic behaviors of ${\rm CO_2}(t)$ and $N(t)$, as required by Eq.~(\ref{CO2-HN}).  To verify the theoretical proposal (\ref{CO2(t)}), we plot the empirical data for ${\rm CO_2}(t)$ on the vertical axis versus ${\rm arccot}[(t_s-t)/\tau_F]$ on the horizontal axis of the parametric Fig.~\ref{Fig:CO2-arccot} using time $t$ as a parameter.  The black data points follow the straight orange line after 1700, when the fitting parameter $\tau_F$ is tuned to the optimal value $\tau_F=40$~years.  Then, the slope and the intersect of the straight line give the values $F_0=116$~ppm and ${\rm CO}_2^{\rm base}=260$~ppm.  Thus, Fig.~\ref{Fig:CO2-arccot} confirms that Eq.~(\ref{CO2(t)}) with the specified parameters, indeed, gives a good fit of the empirical ${\rm CO_2}(t)$ since 1700.

The vertical dashed lines in Fig.~\ref{Fig:CO2-arccot} mark the characteristic phases 0, $\pi/4$, $\pi$, $3\pi/4$, and $\pi$ of the function ${\rm arccot}[(t_s-t)/\tau_F]$ on the horizontal axis, corresponding to the times $t=-\infty$, $t_s-\tau_F$, $t_s$, $t_s+\tau_F$, and $+\infty$.  According to Eq.~(\ref{CO2(t)}), the vertical coordinates for these points are spaced equidistantly with the increment of $F_0\pi/4$, as shown by the dashed horizontal lines and indicated on the right vertical axis of Fig.~\ref{Fig:CO2-arccot}.  The blue labels with ppm units indicate the past and future values of ${\rm CO_2^{fit}}(t)$ at these reference points.  Moreover, these ppm levels of $\rm CO_2$ can be converted into the increments of global temperature in degrees of Celsius using the conversion factor of $100\:{\rm ppm}\approx 1^\circ\:{\rm C}$.\footnote{\url{https://factsonclimate.org/infographics/concentration-warming-relationship}} The projected temperatures increments are shown on the right axis of Fig.~\ref{Fig:CO2-arccot}.  The temperature increase of $1.8^\circ\:{\rm C}$ is predicted by $t_s=2030$, which is consistent with the recognition that the threshold level of $1.5^\circ\:{\rm C}$ has been exceeded \cite{1.5C} (although the definition of ${\rm CO}_2^{\rm base}$ in Fig.~\ref{Fig:CO2-arccot} may be somewhat different from international conventions).  Figure~\ref{Fig:CO2-arccot} also predicts the temperature increase of $2.7^\circ\:{\rm C}$ by $t_s+\tau_F=2070$.  Predictions for more distant future should not be considered reliable.  The projections shown in Fig.~\ref{Fig:CO2-arccot} are based on extrapolation of the past data for ${\rm CO_2}(t)$ and may be disrupted by cataclysmic climate events.  Nevertheless, they may serve as a realistic guide for the near future based on quantitative fitting of the ${\rm CO_2}(t)$ data for the last three centuries from 1700 to the present time.

The red curve in the inset of Fig.~\ref{Fig:CO2-N} shows the theoretical function ${\rm CO_2^{fit}}(t)$ from Eq.~(\ref{CO2(t)}), which agrees well with the ${\rm CO_2}(t)$ data shown by the open blue circles.  The curve has positive (upward) curvature for $t<t_s$ and negative (downward) curvature for $t>t_s$, whereas $t_s=2030$ is the inflection point.  The blue curve at the bottom of the inset in Fig.~\ref{Fig:CO2-N} shows the derivative of the function (\ref{CO2(t)})
\begin{equation}  \label{dCO2/dt}
  \frac{d{\rm CO_2^{fit}}(t)}{dt} = \frac{F_0\tau_F}{(t_s-t)^2+\tau_F^2},
\end{equation}
which represents annual increments in $\rm CO_2$ concentration.  Equation (\ref{dCO2/dt}) is a Lorentzian function with the time width $2\tau_F=80$ years at half-height.  As shown in the inset of Fig.~\ref{Fig:CO2-N}, the annual $\rm CO_2$ increment is expected to peak in the year $t_s=2030$ at the maximal level ${\rm Max}=F_0/\tau_F=2.9$~ppm/year and then decrease to ${\rm Max}/2$ in the year $t_s+\tau_F=2070$.  So far, the global annual $\rm CO_2$ emissions increase every year, but Eq.~(\ref{dCO2/dt}) makes an optimistically realistic prediction that they will peak at $t_s=2030$ and then, 40 years later, will decrease in half.  For comparison, the green curve in the inset of Fig.~\ref{Fig:CO2-N} shows $N_{\rm sqrt}(t)$ predicting a population peak in the same year $t_s=2030$, but of the narrower width $2\tau=64$~years.

The green curve in Fig.~\ref{Fig:QQ} shows a parametric graph of the theoretical predictions for human population $N_{\rm sqrt}(t)$ from Eq.~(\ref{1/N_sqrt-HB}) and for the carbon dioxide level ${\rm CO_2^{fit}}(t)$ from Eq.~(\ref{CO2(t)}).  We observe a good agreement with the empirical data points shown by the black circles.  The green curve exhibits a peak at $t_s=2030$ and a subsequent decrease, because human population is expected to peak, whereas $\rm CO_2$ level continues to increase monotonically.  The green curve reduces to the straight orange line for the time interval 1700--2000 when both $N_{\rm sqrt}(t)$ and ${\rm CO_2^{fit}}(t)$ behave hyperbolically.  The parameters of the straight line can be obtained by using a hyperbolic asymptotic form of Eq.~(\ref{CO2(t)})
\begin{equation}  \label{CO2(t)-hyper} 
  {\rm CO_2^{fit}}(t) \approx {\rm CO}_2^{\rm base} + \frac{F_0 \tau_F}{t_s-t} 
  \qquad {\rm for} \qquad t_s-t \gg \tau_F.
\end{equation}
On the other hand, assuming that $|t_s-t|\ll T_{H\!B}$ and expanding the exponential function in Eq.~(\ref{N_HyperBose}) in the small parameter $|t_s-t|/T_{H\!B}\ll1$,
\begin{equation}  \label{Taylor} 
  e^{(t_s-t)/T_{H\!B}} \approx 1 + \left(\frac{t_s-t}{T_{H\!B}}\right)
  + \frac12 \, \left(\frac{t_s-t}{T_{H\!B}}\right)^2,
\end{equation}
we find
\begin{equation}  \label{N_HyperBose-hyper} 
  N_{\rm HyperBose}(t) \approx N_{H\!B}\left( \frac{T_{H\!B}}{t_s-t}
  - \frac12 \right).
\end{equation}
Comparing Eqs.~(\ref{CO2(t)-hyper}) and (\ref{N_HyperBose-hyper}), we recover the linear relation (\ref{CO2-HN}) with the coefficient
\begin{equation}  \label{K}
  K = \frac{F_0 \tau_F}{C_{H\!B}} = 17.7~\frac{\rm ppm}{\mbox{billion people}},
\end{equation}
where $C_{H\!B}$ is defined in Eq.~(\ref{N_hyp}).  The constant terms in Eqs.~(\ref{CO2-HN}),(\ref{CO2(t)-hyper}), and (\ref{N_HyperBose-hyper}) give the following relation
\begin{equation}  \label{const}
  {\rm CO_2^{offset}} - {\rm CO}_2^{\rm base} = \frac{KN_{H\!B}}2
  = 3.9~{\rm ppm}.
\end{equation}
Using the value ${\rm CO}_2^{\rm base}=260$~ppm, we find that ${\rm CO}_2^{\rm offset}=264$~ppm.  The straight orange line in Fig.~\ref{Fig:QQ} has the parameters $K$ and $\rm CO_2^{offset}$ given by Eqs.~(\ref{K}) and (\ref{const}).

\section{Coupled dynamics of human population and $\rm CO_2$ emissions}
\label{Sec:Dynamics}

This Section gives a mathematical derivation of the anti-Verhulst equation (\ref{N'=N+N2}), which was presented \textit{ad hoc} in Sec.~\ref{Sec:Bose}.  In the spirit of evolutionary biology, let us introduce the fitness function $F(t)$, which determines the growth rate of population $r(t)=F(t)/C_F$ up to a normalizing coefficient $C_F$.  Broadly speaking, the fitness function $F(t)$ represents the achievements of human civilization, such as knowledge, technology, infrastructure, standards of living, and so on, which permit humans to out-compete the rest of the biosphere and thus enable the spectacular growth of human population.  Importantly, the variable $F(t)$ is not constant, but grows in time proportionally to itself (e.g.,\ proportionally to the already accumulated knowledge) and proportionally to human population $N(t)$ (which continues to expand knowledge).  Thus we obtain two coupled ordinary differential equations
\begin{eqnarray}
  && \frac{dN(t)}{dt} = r(t)\,N(t) = \frac{F(t)}{C_F}\,N(t), 
  \label{N'=FN} \\
  && \frac{dF(t)}{dt} = \frac{N(t)}{C}\,F(t).
  \label{F'=FN}
\end{eqnarray}
Subtracting Eq.~(\ref{N'=FN}) from Eq.~(\ref{F'=FN}) with appropriate coefficients, we find that a linear combination of $F(t)$ and $N(t)$ remains constant in time
\begin{equation}  \label{F-N} 
  \frac{d}{dt} \left[ \frac{F(t)}{C_F} - \frac{N(t)}{C} \right] = 0
  \qquad \Longrightarrow \qquad
  \frac{F(t)}{C_F} - \frac{N(t)}{C} = {\rm const}  = \frac{1}{T}.
\end{equation}
Thus, the rate of growth $r(t)$ is a sum of a constant term and a term proportional to population $N(t)$: 
\begin{equation}  \label{r(t)} 
  r(t) = \frac{F(t)}{C_F} = \frac{1}{T} + \frac{N(t)}{C} = \frac{N_B+N(t)}{C}.
\end{equation}
Substituting Eq.~(\ref{r(t)}) into Eq.~(\ref{N'=FN}), we obtain the anti-Verhulst equation (\ref{N'=N+N2}).  

The anti-Verhulst equation (\ref{N'=N+N2}) has been derived in the literature (without using such name) in several ways.  Cohen \cite{Cohen-1995} proposed two coupled Malthus-Condorcet equations for human population growth, which reduce to either the Verhulst equation (\ref{N'=N-N2}) or the anti-Verhulst equation (\ref{N'=N+N2}) depending on the value of a parameter.  For the latter case, he obtained the Bose function (\ref{N_Bose}) in footnote (35) of his paper \cite{Cohen-1995} and pointed out that it possesses a finite-time singularity.  To derive his Malthus-Condorcet equations, Cohen argued that the carrying capacity increases with increasing population.  However, we consider the term ``carrying capacity'' to be mathematically meaningless for the anti-Verhulst equation (\ref{N'=N+N2}).  Instead, our narrative argues that the growth rate $r(t)$ increases with population in Eq.~(\ref{r(t)}), without invoking the concept of carrying capacity.  Sibani and Rasmussen \cite{Sibani-2020} also proposed an equation similar to Eq.~(\ref{r(t)}) and thus obtained the anti-Verhulst equation (\ref{N'=N+N2}) and the Bose function (\ref{N_Bose}).  Positive dependence of the growth rate $r(t)$ on the population size $N(t)$ in Eq.~(\ref{r(t)}) is reminiscent of the Allee effect\footnote{\url{https://en.wikipedia.org/wiki/Allee_effect}} in biology.  From a different perspective, Corsi and Sornette \cite{Corsi-2014} investigated speculative bubbles in financial markets.  In their Appendix A, they studied the same differential equations as (\ref{N'=FN}) and (\ref{F'=FN}), but with the variable $F(t)$ representing the price of an asset and $N(t)$ the amount of money invested in this asset.  Consequently, Corsi and Sornette \cite{Corsi-2014} obtained the conservation law (\ref{F-N}), the anti-Verhulst equation (\ref{N'=N+N2}), and the Bose function (\ref{N_Bose}).  In conclusion, despite differences in language and terminology, all of these narratives produce the same mathematical equation (\ref{N'=N+N2}) for self-reinforced growth resulting in a finite-time singularity.

We tacitly assumed that the constant $1/T$ in Eq.~(\ref{F-N}) is positive.  Taking this constant to be negative produces an equation proposed for city growth by Bettencourt \textit{et al.}\ \cite{Bettencourt-2007}.  It is similar to Eq.~(\ref{N'=N+N2}), but with a negative sign on the first term linear in $N$ and with the second term $N^2$ having been replaced by $N^\beta$.  Depending on whether the initial value of $N$ is below or above a threshold, this equation results in either a decrease $dN/dt<0$ and a collapse, or explosive growth $dN/dt>0$ and a finite-time singularity for $\beta>1$.  West \cite{West-2023} applied the same equation to the growth of global human population.  While the threshold behavior may be relevant for cities (small cities collapse, whereas big cities grow explosively), it does not agree with the data on global human population shown in Fig.~\ref{Fig:Log}.

The fitness function $F(t)$ in Eqs.~(\ref{N'=FN}) and (\ref{F'=FN}) enables population growth.  But any kind of growth requires an energy source.  For life on Earth, it is the energy from the Sun, which is captured by plants via photosynthesis and passed on to animals eating the plants.  For thousands of years, the growth of human population was limited by the available live biomass in plants and animals, first in the wild and then cultivated in agriculture.  Yet, with the onset of the Industrial Revolution, humans learned to exploit the seemingly unlimited resource of dead biomass, which had accumulated for millions of years in the form of fossil fuels.  This gave humans an extraordinary advantage over other species and enabled the unconstrained, scale-free hyperbolic growth described in Sec.~\ref{Sec:Hyperbolic}.  Effectively, fossil fuels became the primary energy source for human growth after the Industrial Revolution until recently, at least.  Thus, we propose to associate the fitness function $F(t)$ after 1700 with the excessive atmospheric carbon dioxide concentration ${\rm CO_2}(t)$ due to the extraction and burning of fossil fuels by humans:
\begin{equation}  \label{F->CO2} 
  F(t) = \left[ {\rm CO_2}(t)-{\rm CO_2^{base}} \right] 
  + \left[ {\rm CO_2^{offset}} -{\rm CO_2^{base}} \right].
\end{equation}
The first two terms in Eq.~(\ref{F->CO2}) correspond to the excessive $\rm CO_2$ concentration above the base level, and the last two terms represent a constant offset.  Then the empirically found linear relation (\ref{CO2-HN}) between ${\rm CO_2}(t)$ and population $N(t)$ confirms the theoretically derived conservation law (\ref{F-N}).  Using Eqs.~(\ref{CO2-HN}) and (\ref{const}), we find that the fitness function (\ref{F->CO2}) is equivalent to
\begin{equation}  \label{F->N} 
  F(t) = K [N(t)+N_{H\!B}].
\end{equation}
The function (\ref{F->N}) satisfies Eq.~(\ref{r(t)}) for $r(t)$ with $K=C_F/C_{H\!B}$, resulting in the HyperBose equation and the HyperBose solution (\ref{N_HyperBose}), which describes population growth after 1700.  This finding validates our claim that the fitness function $F(t)$ is, indeed, directly related to ${\rm CO_2}(t)$ via Eq.~(\ref{F->CO2}).  Then Eq.~(\ref{N'=FN}) indicates that the extraction of fossil fuels enables population growth, whereas Eq.~(\ref{F'=FN}) indicates that population growth enables the extraction of fossil fuels.  These two nonlinear coupled processes mutually enhance one another, thus producing the explosive hyperbolic growth discussed in Sec.~\ref{Sec:Hyperbolic}.

\section{Coupled dynamics in terms of a complex variable}
\label{Sec:Complex}

The coupled equations (\ref{N'=FN}) and (\ref{F'=FN}) possess a finite-time singularity, but do not describe how it becomes truncated.  In this Section, we present an alternative description of the coupled dynamics of population and $\rm CO_2$ level in terms of a complex variable.  It is motivated by observation that, in physics, a common method for regularizing singularities is to shift them away from the real axis into the complex plane.

Let us introduce a complex variable $\Psi(t)$ that satisfies the hyperbolic equation (\ref{N'=N2}) for real time $t$ with a real parameter $C$
\begin{equation}  \label{Psi'=Psi2} 
  \frac{d\Psi}{dt} = \frac{\Psi^2}{C}.
\end{equation}
The solution of this equation has the same pole singularity as in Eq.~(\ref{N_hyp}), but now the pole position $t_s\to t_s-i\tau$ is shifted into the complex plane, where parameters $t_s$ and $\tau$ are determined by initial conditions:
\begin{equation}  \label{Psi-pole} 
  \Psi(t) = \frac{C}{t_s-i\tau-t}.
\end{equation}
The complex variable $\Psi(t)$ can be represented in terms of amplitude and phase as
\begin{equation}  \label{N-phi} 
  \Psi(t) = N(t) \, e^{i\phi(t)}.
\end{equation}
We interpret the amplitude $|\Psi(t)|$ as population $N(t)$, whereas the phase $\phi(t)$ serves as a ``control variable.''  Substituting Eq.~(\ref{Psi-pole}) into Eq.~(\ref{N-phi}), we find that $N(t)$ reproduces Eq.~(\ref{N_sqrt-L}) for $N_{\rm sqrtL}(t)$, whereas the phase $\phi(t)$ is given by the arccot function:
\begin{eqnarray}
  && N(t) = |\Psi(t)| = \frac{C}{\sqrt{(t_s-t)^2+\tau^2}},
  \label{N} \\
  && \phi(t) = {\rm arccot}\left(\frac{t_s-t}{\tau}\right).
  \label{phi}
\end{eqnarray}
Thus, the approach in terms of a complex variable gives an alternative derivation of the cutoff prescribed by $N_{\rm sqrtL}(t)$, complementing the avoided-crossing concept described in Sec.~\ref{Sec:Peak}.  On the other hand, Eq.~(\ref{phi}) is similar to Eq.~(\ref{CO2(t)}).  Thus, it is tempting to associate the phase $\phi(t)$ of the complex variable $\Psi(t)$ with the excessive $\rm CO_2$ level above the baseline:
\begin{equation}  \label{phi-CO2} 
  \phi(t) = \frac{\rm CO_2^{fit}(t) - {\rm CO}_2^{\rm base}}{F_0}.
\end{equation}
However, the simultaneous mapping of $N(t)$ to $|\Psi(t)|$ and ${\rm CO_2}(t)$ to $\phi(t)$ is not quite perfect because of a difference in the fitted time width $\tau=32$ years for the former and $\tau_F=40$ years for the latter.  But the difference is only about 20\%, so this correspondence can be taken as approximate.

It is instructive to rewrite Eq.~(\ref{Psi'=Psi2}) in terms of $N(t)$ and $\phi(t)$.  Substituting Eq.~(\ref{N-phi}) into Eq.~(\ref{Psi'=Psi2}) and separating real and imaginary parts, we obtain two coupled equations
\begin{eqnarray}
  && \frac{dN(t)}{dt} = \frac{N^2(t) \cos\phi(t)}{C},
  \label{cos} \\
  && \frac{d\phi(t)}{dt} = \frac{N(t) \sin\phi(t)}{C}.
  \label{sin}
\end{eqnarray}
Let us start from the initial conditions where $N\neq0$ and $\phi=0^+$, i.e., $\phi$ is positive and close to zero.  Then, $\cos\phi\approx 1$ and Eq.~(\ref{cos}) reduces to Eq.~(\ref{N'=N2}).  However, as $N(t)$ increases, the phase $\phi(t)$ also increases due to Eq.~(\ref{sin}).  When $\phi(t)$ reaches the value $\pi/2$ at $t=t_s$, $\cos\phi(t)$ vanishes in Eq.~(\ref{cos}) and then changes sign for $t>t_s$.  Thus, the population $N(t)$ reaches a peak at $t=t_s$ and then begins to decrease for $t>t_s$.

We see that the phase $\phi(t)$ acts a control variable.  Depending on its value, $0<\phi<\pi/2$ or $\pi/2<\phi<\pi$, we have either $\cos\phi>0$ or $\cos\phi<0$ in Eq.~(\ref{cos}).  In these two regimes, the nonlinear interaction among humans proportional to $N^2$ results in either increasing, $dN/dt>0$, or decreasing, $dN/dt<0$, population, with a transition at $t=t_s$ where $\phi(t_s)=\pi/2$.  This behavior is qualitatively consistent with the concept of a demographic transition.\footnote{\url{https://en.wikipedia.org/wiki/Demographic_transition}}  A demographic transition has two time markers: the beginning, where population starts to grow due to a reduction in the death rate $d$, and the end, where population stabilizes or starts to decrease due to a reduction in the birth rate $b$.  For global population, we identify the beginning with the crossover time $t_c\approx1693$ (shown in Figs.~\ref{Fig:Log} and \ref{Fig:1/N}) of the transition to the fast hyperbolic growth due to massive use of fossil fuels, whereas the end corresponds to the peak at $t_s=2030$.

It is straightforward to transform the more general HyperBose equation (\ref{N'=N+N2}) into a complex form:
\begin{equation}  \label{Psi'=Psi+Psi2} 
  \frac{d\Psi}{dt} = \frac{\Psi}{T_{H\!B}} + \frac{\Psi^2}{C_{H\!B}}.
\end{equation}
Substituting Eq.~(\ref{N-phi}) into Eq.~(\ref{Psi'=Psi+Psi2}) and separating real and imaginary parts, we find
\begin{eqnarray}
  && \frac{dN}{dt} = \frac{N}{T_{H\!B}} + \frac{N^2\cos\phi(t)}{C_{H\!B}} ,
  \label{cos'} \\
  && \frac{d\phi(t)}{dt} = \frac{N(t) \sin\phi(t)}{C_{H\!B}}.
  \label{sin'}
\end{eqnarray}
For $0<\phi<\pi/2$ and $\cos\phi>0$, Eq.~(\ref{cos'}) is similar to the anti-Verhulst equation (\ref{N'=N+N2}).  But when $\phi$ exceeds $\pi/2$, so that $\pi/2<\phi<\pi$ and $\cos\phi<0$, Eq.~(\ref{cos'}) becomes similar the Verhulst equation (\ref{N'=N-N2}).  At $t\to+\infty$, the system arrives to a stable stationary point, where $\phi\to\pi$ and $N\to N_{H\!B}=C_{H\!B}/T_{H\!B}$.  Population stabilizes at the carrying capacity $N_{H\!B}$, but approaches it from above, not from below, after experiencing an overshoot with the peak population $N_{\rm max}$ at $t=t_s$.

\section{Conclusion}
\label{Sec:Conclusion}

Previous studies, such as ``The Limits to Growth'' \cite{Meadows-1972,Bardi-2011}, considered the possibility of a population decline due to the exhaustion of fossil fuels and other resources.  Climate change due to the massive burning of fossil fuels also produces a detrimental effect.  However, our analysis in Sec.~\ref{Sec:Adiabatic} suggests that a decline may happen in the near future simply because the ever-accelerating hyperbolic growth of human population will have reached its natural physical limit, the life-span of a human being, and, thus, cannot accelerate any further.  This makes the present time completely unprecedented in human history.  In addition, while the original goal of our paper was to verify the hyperbolic growth of population $N(t)$ found by von Foerster {\it et al.}~\cite{vonFoerster-1960}, we discovered in the process that the atmospheric carbon dioxide level ${\rm CO_2}(t)$ also follows the same hyperbolic pattern.  Our mathematical fit (\ref{CO2(t)}) predicts the end of this hyperbolic growth with a peak in the annual $\rm CO_2$ emissions at the same year $t_s=2030$, when the population $N(t)$ is projected to reach a peak.

A 21st century peak in human population has been debated by demographers for quite some time \cite{Lutz-2001}.  At long last, the topic of a demographic peak took center stage in the latest UN report of 2024 \cite{UN-2024}, where a search for the string ``peak'' brings 54 entries, compared to only 8 entries in the previous report of 2022 \cite{UN-2022}.  Now, the \textit{first} ``key message'' in the 2024 report \cite{UN-2024} forecasts a peak in the world's population in the mid-2080s.  A similar message was the 7th item on the key list in the 2022 report \cite{UN-2022}, supplemented by a contradictory caveat that the population is expected ``to remain at that level until 2100,'' which suggests it is not really a peak.  The prediction by the UN of a population peak in mid-2080s is based on the ``most probable'' medium scenarios.  In contrast, from our mathematical fit of $N(t)$, we predict a population peak at the much earlier year $t_s=2030$, which, nevertheless, is roughly compatible with the low scenarios by Lutz {\it et al.}~\cite{Lutz-2001} (see their Fig.~2) and the UN report of 2022 \cite{UN-2022} (see their Fig.~III.4).  As Ref.~\cite{WSJ-2024} points out, from 2017 to 2022, the UN has lowered its forecasts and moved the expected peak for medium scenarios closer to the present.  Further updates are likely to continue moving it in this direction \cite{WSJ-2024}.

Even if the world's population decline is delayed beyond our predicted year 2030 by a decade or two, such a delay would be minuscule on the scale of human history and would not alter the main findings of our paper.  But it is advisable to start preparing for an early peak, rather than wait longer.  Arguably, the question of growth, particularly of human population, is one of the central fundamental challenges of the 21st century \cite{Safa-2016,Gronewold-2021}.  It is intricately intertwined with other significant issues faced by humankind in the context of what is sometimes referred to as a polycrisis \cite{Homer-Dixon-2023}.

\section*{Data Availability}
The input data and the Matlab script for data fitting and generation of the figures are included with submission of this manuscript for public posting at \url{https://doi.org/10.1016/j.physa.2025.130412}

\section*{Acknowledgments}
Ethan Levy was involved in preliminary data analysis during an early stage of this project.  At a later stage, Andrew Dirr performed data collection and analysis, code generation and parameter fitting, and produced figures for the initial version of the paper.  Subsequently, Thomas and Anatoley Zheleznyak revised fitting procedures and produced final versions of all plots in this paper with the most recently available data.



\begin{thebibliography}{99}

\bibitem{vonFoerster-1960} 
  H. von Foerster, P.M. Mora, L.W. Amiot, 
  Doomsday: Friday, 13 November, A.D. 2026,
  Science 132 (1960) 1291--1295.
  \url{https://doi.org/10.1126/science.132.3436.1291}

\bibitem{Umpleby-1990}
  S.A. Umpleby,  
  The scientific revolution in demography, 
  Popul. Environ. 11 (1990) 159--174.
  \url{https://doi.org/10.1007/BF01254115}

\bibitem{WSJ-2024}
  G. Ip, J. Adamy,
  Suddenly there aren’t enough babies: The whole world is alarmed,
  The Wall Street Journal, 13 May 2024 online, 14 May 2024 print. 
  \url{https://www.wsj.com/world/birthrates-global-decline-cause-ddaf8be2}
  (accessed 4 January 2025)

\bibitem{Johansen-2001}
  A. Johansen, D. Sornette,  
  Finite-time singularity in the dynamics of the world population, economic and financial indices, 
  Physica A 294 (2001) 465--502.
  \url{https://doi.org/10.1016/S0378-4371(01)00105-4}

\bibitem{Sornette-2002} 
  D. Sornette, Why Stock Markets Crash,
  Princeton University Press, Princeton, 2002, Ch. 10, ISBN 0-691-09630-9

\bibitem{Safa-2016}
  S. Motesharrei, {\it et al.},
  Modeling sustainability: population, inequality, consumption, and bidirectional coupling of the Earth and Human Systems,  
  Natl. Sci. Rev. 3 (2016) 470--494.
  \url{https://doi.org/10.1093/nsr/nww081}

\bibitem{Sibani-2020}
  P. Sibani, S. Rasmussen,
  Human wealth evolution: trends and fluctuations, 
  Physica A 558 (2020) 124985.
  \url{https://doi.org/10.1016/j.physa.2020.124985}

\bibitem{Gleria-2024}
  I. Gleria, S. DaSilva, L. Brenig, T.M.R. Filho, A. Figueiredo, 
  Modified Verhulst–Solow model for long-term population and economic growth,
  J. Stat. Mech. (2024) 023406.
  \url{https://doi.org/10.1088/1742-5468/ad267a}

\bibitem{Kuretova-2010}
  E.D. Kuretova, E.S. Kurkina, 
  Modeling general laws of spatial-temporal evolution of society: hyperbolic growth and historical cycles,
  Comput. Math. Model. 21 (2010) 70--89.
  \url{https://doi.org/10.1007/s10598-010-9055-9}

\bibitem{Jusup-2022}
  M. Jusup, {\it et al.}, Social physics,  
  Physics Reports 948 (2022) 1--148.
  \url{https://doi.org/10.1016/j.physrep.2021.10.005}

\bibitem{Our-World-in-Data}
  Our World in Data, 2024,
  Population, 10,000 BCE to 2023. 
  \url{https://ourworldindata.org/grapher/population}, Web update 15 July 2024, downloaded 2 January 2025.
    
\bibitem{WorldBank}
  The World Bank, 2024,
  Population, total (up to 2023).
  \url{https://data.worldbank.org/indicator/SP.POP.TOTL}, Web update 16 December 2024, downloaded 2 January 2025.

\bibitem{Monceau-2001}
  P. Monceau, F.Ya. Nad, S. Brazovskii,
  Ferroelectric Mott-Hubbard phase of organic TMTTF$_2$X conductors,
  Phys. Rev. Lett. 86 (2001) 4080--4083.
  \url{https://doi.org/10.1103/PhysRevLett.86.4080}

\bibitem{Kapitza-1996}
  S.P. Kapitza,
  The phenomenological theory of world population growth,
  Physics-Uspekhi 39 (1996) 57--71.
  \url{https://doi.org/10.1070/pu1996v039n01abeh000127}

\bibitem{Kapitza-2013}
  S.P. Kapitza, 2013,
  Scaling in modelling global population growth.
  \url{https://spkurdyumov.ru/uploads/2013/08/Kapitsa400.pdf}  
  (accessed 4 January 2025)

\bibitem{Husler-2014}
  A.D. H\"usler, D. Sornette, 
  Human population and atmospheric carbon dioxide growth dynamics: Diagnostics for the future,
  Eur. Phys. J. Spec. Top. 223 (2014) 2065--2085.
  \url{http://dx.doi.org/10.1140/epjst/e2014-02250-7}

\bibitem{Bongaarts-2018}
  J. Bongaarts, B.C. O’Neill, 
  Global warming policy: Is population left out in the cold?,
  Science 361 (2018) 650--652.
  \url{http://dx.doi.org/10.1126/science.aat8680}

\bibitem{EPA-CO2}
  U. S. Environmental Protection Agency (EPA), 2024,
  Atmospheric Concentrations of Greenhouse Gases.
  \url{https://www.epa.gov/climate-indicators/climate-change-indicators-atmospheric-concentrations-greenhouse-gases}, Data File \url{https://www.epa.gov/system/files/other-files/2024-06/ghg-concentrations_fig-1.csv}, Web update June 2024, downloaded 2 January 2025.

\bibitem{CO2-stay}
  A. Buis,
  The atmosphere: Getting a handle on carbon dioxide (Part Two),
  NASA Science Editorial Team, 9 October 2019 online. 
  \url{https://science.nasa.gov/earth/climate-change/greenhouse-gases/the-atmosphere-getting-a-handle-on-carbon-dioxide/}
  (accessed 4 January 2025)

\bibitem{1.5C}
  O. Milman,
  World’s 1.5C climate target ‘deader than a doornail’, experts say,
  The Guardian, 18 November 2024 online. 
  \url{https://www.theguardian.com/environment/2024/nov/18/climate-crisis-world-temperature-target}
  (accessed 4 January 2025)

\bibitem{Cohen-1995}
  J.E. Cohen, Population growth and Earth’s human carrying capacity,
  Science 269 (1995) 341--346.
  \url{https://doi.org/10.1126/science.7618100}

\bibitem{Corsi-2014}
  F. Corsi, D. Sornette, 
  Follow the money: The monetary roots of bubbles and crashes,
  International Review of Financial Analysis 32 (2014) 47--59.
  \url{http://dx.doi.org/10.1016/j.irfa.2014.01.007}

\bibitem{Bettencourt-2007}
  L.M.A. Bettencourt, J. Lobo, D. Herbing, C. Kuehnert, G.B. West,
  Growth, innovation, scaling, and the pace of life in cities, 
  Proc. Natl. Acad. Sci. USA 104 (2007) 7301--7306.
  \url{https://doi.org/10.1073/pnas.0610172104}

\bibitem{West-2023}
  G. West, On the future of the planet; universal laws of life, 
  growth and death from organisms and cities to the anthroposphere,
  recorded Zoom talk at the Mathematical Physics Webinar, Rutger University, 22 March 2023, 
  \url{https://cmsr.rutgers.edu/news-events-cmsr/mathematical-physics-webinar/range.listevents/-} 
  (accessed 4 January 2025)

\bibitem{Meadows-1972}
  D.H. Meadows, D.L. Meadows, J. Randers, W.W. Behrens III,
  The Limits to Growth, New York, Universe Books, 1972, ISBN 0-87663-165-0
 
\bibitem{Bardi-2011}
  U. Bardi, The Limits to Growth Revisited, 
  Springer, New York, 2011, ISBN 978-1-4419-9415-8.
  \url{https://doi.org/10.1007/978-1-4419-9416-5}

\bibitem{Lutz-2001}
  W. Lutz, W. Sanderson, S. Scherbov,
  The end of world population growth,
  Nature 412 (2001) 543--545.
  \url{https://doi.org/10.1038/35087589}

\bibitem{UN-2024}
  United Nations Department of Economic and Social Affairs: Population Division, 
  World Population Prospects 2024: Summary of Results. 
  \url{https://population.un.org/wpp/assets/Files/WPP2024_Summary-of-Results.pdf} 
  (accessed 4 January 2025)
  
\bibitem{UN-2022}
  United Nations Department of Economic and Social Affairs: Population Division, 
  World Population Prospects 2022: Summary of Results. 
  \url{https://www.un.org/development/desa/pd/sites/www.un.org.development.desa.pd/files/wpp2022_summary_of_results.pdf} 
  (accessed 4 January 2025)

\bibitem{Gronewold-2021} 
  N. Gronewold, Anthill Economics: Animal Ecosystems and the Human Economy,
  Prometheus Books, Guilford, Connecticut, 2021, ISBN 978-1-633-88653-7

\bibitem{Homer-Dixon-2023}
  T. Homer-Dixon, Why so much is going wrong at the same time, 
  Vox (online), 18 October 2023.
  \url{https://www.vox.com/future-perfect/23920997/polycrisis-climate-pandemic-population-connectivity} 
  (accessed 4 January 2025)


\end{thebibliography}
\end{document}